\documentclass[aps,pra,reprint,superscriptaddress]{revtex4-2}
\usepackage{amsmath,amsfonts}
\usepackage{graphicx}

\begin{document}

\title{Propagation of space-time optical vortices in multimode fibers}

\author{Spencer W. Jolly}
\email{spencer.jolly@ulb.be}
\affiliation{Service OPERA-Photonique, Université libre de Bruxelles (ULB), Brussels, Belgium}
\author{Julien Dechanxhe}
\affiliation{Service OPERA-Photonique, Université libre de Bruxelles (ULB), Brussels, Belgium}
\author{Pascal Kockaert}
\affiliation{Service OPERA-Photonique, Université libre de Bruxelles (ULB), Brussels, Belgium}

\date{\today}

\begin{abstract}
The non-intuitive spatiotemporal modal content of space-time optical vortices (STOVs) is calculated in a graded-index fiber supporting a large number of propagating modes. We discuss how a fiber supporting many modes allows to truly couple higher-order STOVs, the number of modes necessary to support a STOV of a certain order, and conversely the truncation effect in a few-mode fiber. Based on the excited modes and their temporal profiles, we show numerical results for the linear and nonlinear propagation of STOVs in multimode fibers, specifically the linear space-time beating at short propagation distances, and the nonlinear trapping effect between modes producing stable states on long propagation distances. Our results underline how STOVs present a rich platform for multimode nonlinear optics and technology.
\end{abstract}

\maketitle

\section{introduction}

Ultrafast optics in multimode fibers, especially nonlinear propagation and the related effects, is recently experiencing a renaissance enabled by more powerful simulation algorithms and accessible platforms~\cite{krupa19}, and resulting in both new fundamental pictures of multimode interactions~\cite{wright22-2} and progress towards enabled technologies~\cite{wright22-1}. In parallel, space-time optics~\cite{shenY23}, a subset of structured light where the unseparable structure is between the space and time dimensions, has been advancing toward fully arbitrary non-separable light fields~\cite{shenY22}, topological structures, etc. Space-time optical vortices (STOVs)~\cite{wanC23,bekshaev2024spatiotemporal,porras25-1} are a specific example of space-time fields with a vortex singularity in space-time, which have been recently demonstrated in a number of ways~\cite{jhajj16,hancock19,chong20,porras24-1,zhanQ24} and are becoming increasingly interesting for applications such as particle manipulation or sensing, and even high-field phenomena such as particle acceleration or high-harmonic generation. Recently these fields of study are being explored more closely together, whereby space-time optics are starting to be explored more generally and arbitrarily in multimode fibers~\cite{jolly23-2,dechanxhe2025accessingdifferenthigherordermodes,suX25} in the interest of pursuing the information transfer, sensing, imaging, and computing applications and technologies that can be enabled by multimode fibers and guided-wave optics in general.

In this work we will consider how a STOV couples to and propagates within a multimode fiber, focusing first specifically on the modal distribution at the input facet. This modal picture is very instructive, since each mode has its own propagation constants (refractive index, group index, chromatic dispersion, etc.) that fully determine how they propagate in the linear regime (i.e. when the STOV pulse energy is low). We will then show how that modal distribution, where the different modes have different initial phases and temporal profiles, results in important linear propagation effects on short and long length scales. Finally, we will investigate the nonlinear propagation of a STOV and the great potential for fundamental and applied studies.

The most simple way to write the complex electric field of a Gaussian STOV is, ignoring propagation (i.e. at a waist), as follows

\begin{equation}
E=\left(\frac{x}{w_0}\pm\frac{it}{t_0}\right)^n e^{-t^2/t_0^2}e^{-(x^2+y^2)/w_0^2} e^{-i\omega_0 t}.\label{eq:STOV}
\end{equation}

\begin{figure}[t]
\centering
 \includegraphics[width=84mm]{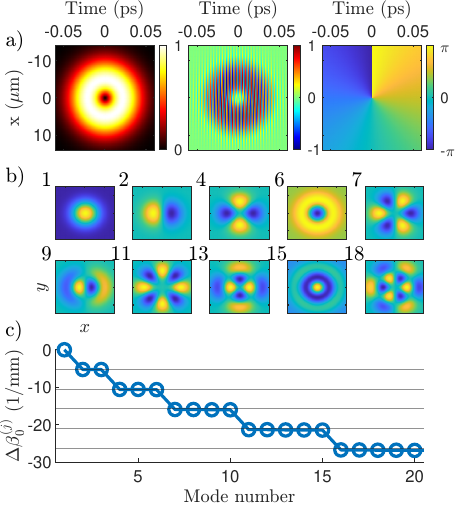}
\caption{A STOV of charge 1, waist $w_0=7\,\mu$m, and duration $t_0=30$\,fs (a) shown with amplitude (left), real field (center), and phase (right), at $y=0$. The first 10 spatial modes of our canonical GRIN fiber (b) that are symmetric in $y$ shown with their mode numbers, with $\{x,y\}\in[-15,15]\,\mu$m. The lowest-order propagation constant $\Delta\beta_0^{(j)}$ (c) for mode number $j$ showing the mode grouping on multiples of $\Delta\beta_0^{(2)}$ (faint lines).} 
\label{fig:1}
\end{figure}

\noindent This STOV, shown in Fig.~\ref{fig:1}(a), is assumed to be elliptical (i.e. $x$ and $y$ have the same size $w_0$ in the Gaussian envelope, and $w_0$ is also the scale length of the vortex term), with a duration $t_0$, and it is of topological charge $l$, where $n=|l|$ and the $\pm$ refers to the sign of $l$. This form of STOV is the most simple to consider mathematically and the most intuitive, and also has the lowest number of free parameters such that it allows for a good starting point when considering the interactions with physical systems or optical devices. In Fig.~\ref{fig:1}(a) we see the characteristics of the STOV, it's donut profile and phase winding in the $x-t$ plane, for a relatively short duration of 30\,fs such that we can see the field oscillations.

We will consider first the coupling of a STOV to and propagation within a parabolic graded-index (GRIN) fiber, which is interesting because it supports spatial mode groups with more closely-spaced propagation constants than the more common step-index fiber. This allows for more rich and long-lasting interactions between modes in nonlinear propagation, and uniform inter-modal beating in linear propagation. We will consider only one GRIN fiber with fixed properties such that the coupling and propagation that we will explore can be tuned only via the properties of the STOV. This silica fiber has a parabolic index profile, with a 25\,{\textmu}m core radius and an index of difference of 0.0137 between the center of the core and the cladding. In Fig.~\ref{fig:1}(b) we show the first 10 modes (and their corresponding numbering according to decreasing propagation constant, shown also in Fig.~\ref{fig:1}(c)) that are important in this work. Notably we will consider only cases that are symmetric along the $y$-axis, since the STOV in Eq.~\ref{eq:STOV} is always Gaussian along $y$ and we will not consider the case of a spatial offset along $y$. This means, for example, that in many mode groups we will only consider half of the modes, the mode asymmetric in $x$ and symmetric in $y$. For example, we consider modes 2, 4, 7, and 9 shown in Fig.~\ref{fig:1}(b), but not the rotated paired modes 3, 5, 8, 10, etc. Cylindrically-symmetric modes (for example modes 1, 6, 15, etc.) play an important role that will be discussed later.

\section{coupling a STOV to a multimode fiber}

Previous works have investigated the coupling of such elliptical STOVs of order 1 and 2 into few-mode fibers, and their following linear propagation, including dispersion~\cite{caoQ23,zhangC24}. These works already exposed significant interesting phenomena when coupling into multimode fibers and the following effects of chromatic and modal dispersion on the intensity and phase distribution of the STOVs. However, as we will show here, considering higher-order STOVs, fibers supporting more modes, and other fiber geometries allows for more complex and potentially rich phenomena.

\begin{figure}[t]
\centering
\includegraphics[width=84mm]{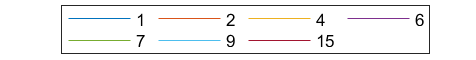}
\includegraphics[width=84mm]{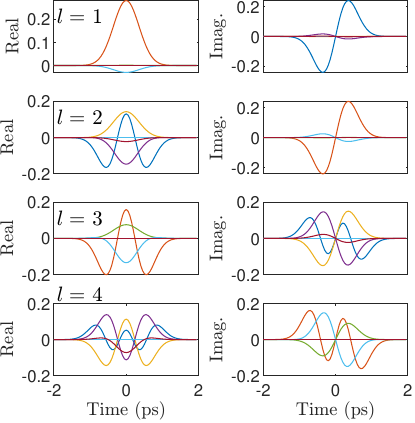}
\includegraphics[width=84mm]{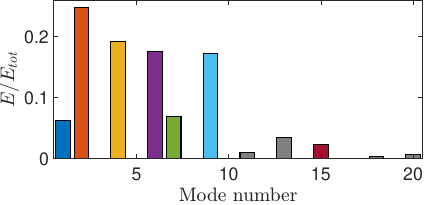}
\caption{Temporal envelopes coupled into the GRIN fiber for 7 example modes (numbered in the legend). The real part (left) and imaginary part (right) of the complex coupling envelopes are shown for STOVs with $l=1$, 2, 3, and 4 (top to bottom) for fixed $w_0=7$\,{\textmu}m and $t_0=500$\,fs. The bottom panel shows the total energy in each mode for $l=4$.} 
\label{fig:2}
\end{figure}

Calculating the coupling into the different modes of our example fiber requires not only modal coefficients, but complex temporal envelopes for each mode using the same strategy as in Refs.~\cite{caoQ23,zhangC24}. Note that in some of our own past work we calculated frequency-dependent complex modal envelopes~\cite{jolly23-2}, but since the STOVs can be described easily directly in time, we can also create the modal envelopes directly in time. We show the resulting mode coupling in Figure~\ref{fig:2}, for topological charges $l=1$, 2, 3, and 4 for a Gaussian elliptical STOV with a waist $w_0=7$\,{\textmu}m and a duration $t_0=500$\,fs. The waist is chosen since it provides the most pure mode coupling, where for example a Gaussian of $w_0=7$\,{\textmu}m couples almost exclusively to mode 1. This way we can investigate the coupling and subsequent propagation of the STOV with the least number of modes and can be sure that the modal content is due purely to the STOV topology and not a size mismatch with the fiber modes. We show in Fig.~\ref{fig:2} the temporal envelopes (real and imaginary components) for a selection of the most important modes.

The modal content can be further understood by looking purely at the vortex term for different topological charges, and making approximations to the mode profiles. For example, if we approximate the spatial profile of mode 1 ($M_1$) to be a Gaussian~\cite{marcuse78} and the spatial profile of mode 2 ($M_2$) to be a Hermite-Gaussian profile along $x$, then looking at the case of $n=1$ the temporal envelopes can be easily deduced

\begin{align}
\begin{split}
\left(\frac{x}{w_0}\pm\frac{it}{t_0}\right) & e^{-t^2/t_0^2}e^{-(x^2+y^2)/w_0^2}\\
&\rightarrow\quad e^{-t^2/t_0^2}\left[\left(\frac{\pm it}{t_0}\right)M_1 + M_2\right].\label{eq:sketch1}
\end{split}
\end{align}

\noindent In fact the approximations for the spatial profiles $M_1$ and $M_2$ are actually very accurate when the waist is chosen properly, such that this sketch accounts for the vast majority of the coupled energy in the case of $l=\pm 1$, which can be seen in the low amount of energy in other modes in Fig.~\ref{fig:2}. This is why the past work in that case~\cite{caoQ23} remains fully valid even in the case of fibers that support more modes.

Looking at $n=2$, it understandably becomes more complex. Expanding the vortex term in that case and including more spatial mode profiles $M_i$ we can see

\begin{align}
\begin{split}
&\left(\frac{x}{w_0}\pm\frac{it}{t_0}\right)^2 e^{-t^2/t_0^2} e^{-(x^2+y^2)/w_0^2}\\
&\rightarrow\quad e^{-t^2/t_0^2}\Bigg[M_1\left(c_1-\frac{t^2}{t_0^2}\right) \pm \left(\frac{2it}{t_0}\right)M_2 + c_2 M_4\\
&\quad\quad\quad\quad\quad+M_6\left(c_3+c_4\frac{t^2}{t_0^2}\right) \Bigg],\label{eq:sketch2}
\end{split}
\end{align}

\noindent that, although it is less close to the exact results (there are undefined constants $c_1$, $c_2$, etc.), still shows the complexity. We see that mode 1 has a term of the form $(it/t_0)^n$, and that with increasing $n$ we need to include higher order modes. This trend is confirmed in the numerical calculations for $l=3$ and 4 as well, shown in lower rows of Figure~\ref{fig:2}. Finally, the last panel of Fig.~\ref{fig:2} shows an alternative viewpoint, which is the energy in each mode 1--20 for the case of $l=4$. If higher-order modes aren't guided in the fiber, then the guided energy will be truncated and the spatiotemporal field that is guided will not be the same as the STOV in free space.

\begin{figure}[t]
\centering
\includegraphics[width=21mm]{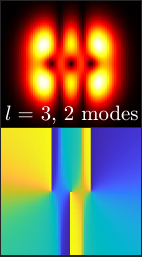}%
\includegraphics[width=21mm]{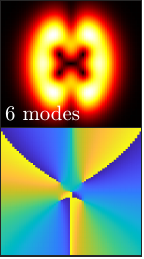}%
\includegraphics[width=21mm]{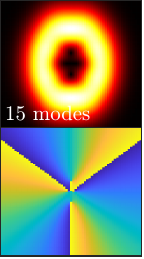}%
\includegraphics[width=21mm]{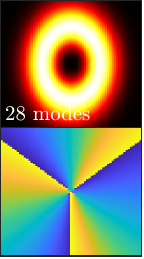}
\caption{Field Amplitude (top of each panel) and phase (bottom of each panel) constructed at the input of the fiber considering only a limited number of modes---simulating a fiber that supports that amount of modes. For fixed $w_0=7$\,{\textmu}m, the field is shown at $y=0$ for $l=3$ for an increasing number of modes considered (left to right). The horizontal axis is $t\in[-3t_0,3t_0]$ and the vertical axis is $x\in[-15,15]\,\mu$m for each panel. The amplitude is normalized in each panel.} 
\label{fig:3}
\end{figure}

In the recent work considering $l=2$ in a fiber containing only a few modes~\cite{zhangC24}, the coupled space-time profile was indeed significantly different than the free-space STOV for the reasons stated above. However, in a fiber supporting more propagating modes as is the case here, our results show that the charge 2 and above singularities can be retained at the input of the fiber. We artificially recreate this scenario in Fig.~\ref{fig:3} by showing the space-time fields reconstructed in the fiber while considering only a certain (increasing) number of modes. In combination with the previous simple sketch, this exercise underscores the importance of the cylindrically-symmetric modes (1, 6, 15, 28, etc.) for resolving the singularity at $x=t=0$. Mode 1 has a temporal profile $\propto(it/t_0)^n$, but to resolve the central singularity we need more than $n$ cylindricaly-symmetric modes such that they interfere to create zero intensity at the center of the STOV. The other modes, asymmetric in $x$, are responsible for increasingly resolving both the donut profile and the uniform cyclical nature of the phase around the singularity. One can conclude, for example, that even in a fiber supporting tens of modes, the coupling of a significantly higher-order STOV would be incomplete even before any propagation in the fiber. We do not show the profile along $y$ in Fig.~\ref{fig:3}, but with insufficient modes in the fiber this would also be complicated and non-Gaussian, although remaining symmetric in $y$. If the waist is not chose to match the size of the fundamental mode of the fiber, then even more modes would be excited, or conversely, it would take more modes to fully guide the STOV structure.

\section{linear propagation}

\begin{figure}[t]
\centering
\includegraphics[width=84mm]{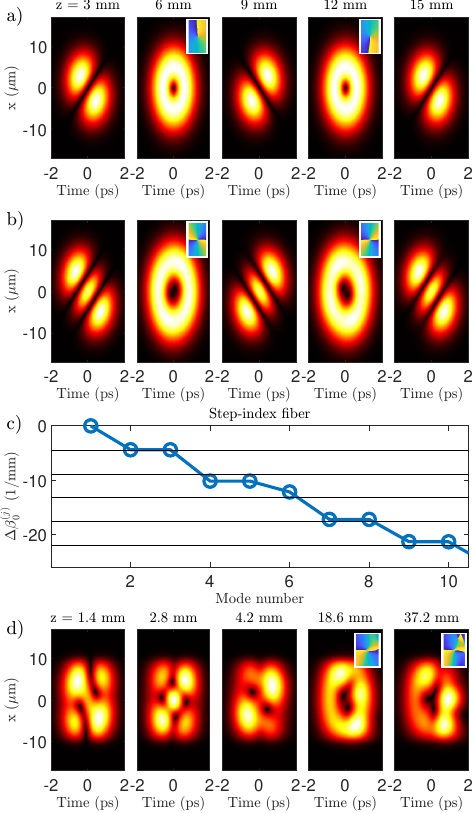}
\caption{Linear propagation of STOVs in multimode fibers. For $l=1$ (a) and $l=2$ (b) in the GRIN fiber, we see almost perfect cyclical reoccurence of the STOV structure at multiples of $L_\textrm{beat}=1.2$\,mm. However, for a step-index fiber of radius 10\,$\mu$m, the phase parameters (c) are not multiples $\Delta\beta_0^{(2)}$ (horizontal lines), meaning that the linear propagation in that fiber (d) for $l=2$ produces no reoccurence, even when the distance is a chosen multiple of 1.43\,mm, the beat length in that case. The amplitude is normalized in each panel.} 
\label{fig:4}
\end{figure}

With short-distance linear propagation, ignoring dispersion, the modes involved will beat with each other due to their different lowest-order propagation constant $\Delta\beta_0^{(j)}=\beta_0^{(j)}-\beta_0^{(1)}$, written relative to the fundamental mode 1 for higher-order mode $j$. This means very simply that the STOV structure and singularity will not be maintained even on very short distances. Looking at $l=1$, and in our simple sketch Eq.~\ref{eq:sketch1}, at a propagation distance $L$ the field of the input STOV becomes approximately $e^{-t^2/t_0^2}\left[\left(\frac{\pm it}{t_0}\right)M_1 + \left(e^{i\Delta\beta_0^{(2)}L}\right)M_2\right]$---after a certain distance the STOV becomes a two-lobed structure $\sim(x/w_0+t/t_0)$, then it becomes a STOV of opposite topological charge, then an opposite two-lobed structure, and then again like the original STOV. This is a perfectly cyclic phenomenon when ignoring dispersion. For the fiber that we consider when pumped at a central wavelength of 1030\,nm, the period of this cyclical STOV inversion and re-appearance is $L_\textrm{beat}=2\pi/|\Delta\beta_0^{(2)}|=1.2$\,mm, which is shown for that case in Fig.~\ref{fig:4}(a) for a pulse of 500\,fs duration. The reversal of the topological charge after $L_\textrm{beat}/2$ calls into question what is occurring in terms of the transverse orbital angular momentum of the STOV, but this is beyond the scope of this work.

These beating dynamics may not be valid, however, for higher-order STOVs since there could be multiple propagation constants involved that are not multiples of each other. This becomes apparent when looking at our earlier sketch for $l=2$ in Eq.~\ref{eq:sketch2}, whereby the linear in time mode 2, constant mode 4, and mode 6 will de-phase differently from mode 1. Considering our specific GRIN fiber, in fact $\Delta\beta_0^{(6)}\approx\Delta\beta_0^{(4)}\approx2\Delta\beta_0^{(2)}$ as seen in the bottom of Fig.~\ref{fig:1}---a well-known advantage of parabolic GRIN fibers. This means that for $l=2$ we can also recover the same STOV as at the input after the same distance as for $l=1$, i.e. $L_\textrm{beat}=1.2$\,mm, which is also shown in Fig.~\ref{fig:4}(b). For orders higher than $l=2$ this remains true, but the dynamics within that characteristic distance become more complex and more quickly oscillating due to the increasingly different propagation constants of the higher-order modes that are excited. Interestingly, the multi-lobed space-time profile at distances within the beating cycle looks strikingly similar to that of a higher-order STOV after it has propagated in free space~\cite{porras23-1}, which manifests in that scenario due to the difference in Gouy phase between a Gaussian and Hermite Gaussian of order $l$ of $|l|\pi/2$. Interestingly this means that, if one took a large collimated STOV and focused it onto the facet of a GRIN fiber such that a space-time lobed structure was coupled, it would turn into a STOV-like structure after propagating $L_\textrm{beat}/4$ and would evolve after that point as if a STOV was coupled at the input facet.

However, for a step-index fiber with a 10\,$\mu$m radius and the same index difference of 0.0137 (producing mode sizes and shapes similar to the GRIN fiber shown in Fig.~\ref{fig:1}), the beta parameters are not generally multiples of each other, as shown also in Fig.~\ref{fig:4}(c). This means that for a STOV $l=2$ there is no simple periodicity of the dephasing between the modes. For example, neither $\Delta\beta_0^{(4)}$ nor $\Delta\beta_0^{(6)}$ are a multiple of $\Delta\beta_0^{(2)}=2\pi/1.43$\,mm$^{-1}$ for the step-index case. If we approximate that the modal content in the $l=2$ case is purely in those modes 1, 2, 4, and 6, then we can only hope to at some distance be \textit{close} to periodic and recover the STOV structure in space-time. In the case of step index fiber with parameters as in Fig.~\ref{fig:4}(c), one can only get relatively close to being in phase for all of modes 1, 2, 4, and 6 at distances, for example, of $13\times2\pi/\Delta\beta_0^{(2)}=18.6$\,mm or $99\times2\pi/\Delta\beta_0^{(2)}=141.5$\,mm---distances chosen because they minimize the average phase difference  between modes 1, 2, 4, and 6. In the former case this is still not exactly periodic, meaning that the STOV is deformed somehow even at those points of minimal dephasing as shown in Fig.~\ref{fig:4}(d). For the latter case, it is long enough that modal (and chromatic) dispersion is significant and the vortex is broken up for that reason. This demonstrates quite a strong conclusion, which is that a STOV of $n>1$ cannot be maintained perfectly over any distance nor even be periodic within a fiber that does not have the $\Delta\beta_0^{(j)}$ parameters grouped as for the GRIN fiber.

\begin{figure*}[t]
\centering
\includegraphics[width=168mm]{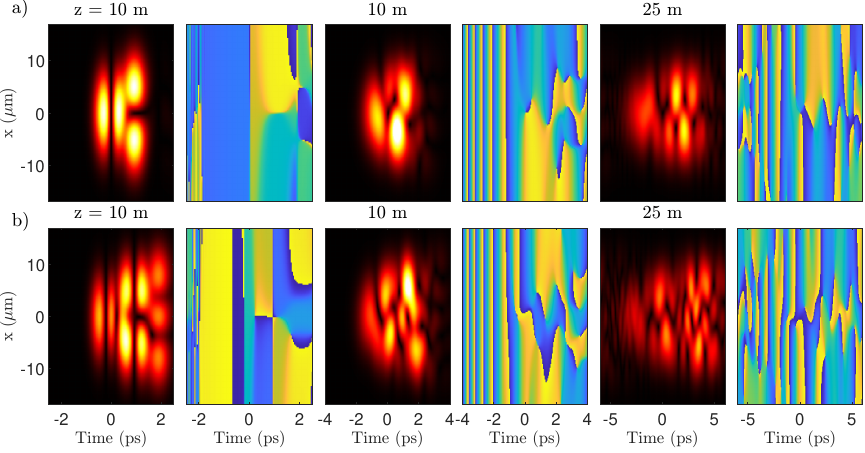}
\caption{Linear propagation of a 500 fs STOV $l=1$ (a) and $l=2$ (b) over different propagation distances in a GRIN fiber. For each propagation distance the field amplitude and its phase are plotted. The left column corresponds to the case of zero dispersion regime ($\lambda_0 = 1270$\,nm), while the central and right columns correspond to the case of a normal dispersion regime ($\lambda_0 = 1030$\,nm).}
\label{fig:5}
\end{figure*}

Upon longer-distance propagation dispersion starts to become significant. In the case of multimode fibers there is both modal dispersion, and standard chromatic dispersion (depending on the central wavelength). Essentially, as corroborated in the past work~\cite{caoQ23,zhangC24}, the STOV structure will breakup mainly due to the modal dispersion, as seen in Fig.~\ref{fig:5}(left) at 1270\,nm---the wavelength of zero chromatic dispersion in our GRIN fiber---for 10\,m of propagation. Eventually due to their different $\beta_1$ parameters (group index) the modes are fully separate in time, and no longer constructively add regardless of their relative phase. For 1270\,nm with $l=1$ in Fig.~\ref{fig:5}(a) we can see modes 1 and 2 are temporally separate, and for $l=2$ in Fig.~\ref{fig:5}(b) we see modes 1, 2, 4, and 6 (although modes 4 and 6 are fully overlapping due to their nearly identical group index). At wavelengths above or below that, at 1030\,nm for example as in Fig.~\ref{fig:5}(center), the increase of the pulse duration allows for vortex-like singularities to persist since the pulse duration of each mode is stretched to greater than the inter-modal delay (i.e. the modes still overlap temporally). Although vortices persist, the donut profile no longer does and the intensity and phase distribution becomes more complex, and increasingly complex as the topological charge $l$ increases.

At even longer propagation distances with 1030\,nm, for example 25\,m as in Fig.~\ref{fig:5}(right), dispersion understandably has a much stronger effect. With $l=1$ there is still some visible beating between mode 1 and mode 2 that creates a specific structure, but with $l=2$ there is nothing discernible besides a more complicated interference between the more than 4 modes present.

As has been discussed in bulk dispersive materials, the dispersion could be tuned arbitrarily to have advantageous linear propagation behavior~\cite{hyde23}. Or, in free-space generation, the STOV could be pre-chirped to control the position where the STOV is eventually formed~\cite{novikov25}. But in bulk media the dispersion applies to the entire spatiospectral profile, i.e. is not separate for the modes. Similar concepts hold true in the case of multimode fibers~\cite{caoQ23}, where pre-chirping and pre-shaping can allow for the STOV to appear at the fiber exit---to a different degree depending on the flexibility of the compensation and the length of propagation. However, the fiber type and geometry strictly control the propagation parameters of the different modes, and especially controlling the phase between modes (related to the $\Delta\beta_0^{(j)}$ parameters) requires precise prior knowledge about the fiber length. This has been discussed more recently in the context of more general toroidal beam structures, using a transmission matrix technique to find the input space-time distribution required to have the desired output distribution~\cite{komonen2025highdimensionalspatiotemporaltoroidal}, but this requires total arbitrary control over the input field and thus a complex apparatus.

\section{nonlinear propagation}

We expect upon nonlinear propagation of a STOV coupled in the same GRIN fiber, significant modal interactions and modification of the instantaneous refractive index can lead to a qualitative change in behavior, potentially allowing for longer persistence of higher-quality STOV beams or the creation of more interesting space-time profiles. However, the fact that different modes are purely real or purely imaginary, as seen in Fig.~\ref{fig:2} (i.e. with a $\pi/2$ phase shift) means that the phase of the nonlinear interactions will be important.

We use the Generalized Multimode Nonlinear Schrödinger Equation (GMMNLSE) and an open-source code to perform nonlinear propagation simulations~\cite{wright18,github_GMMNLSE}. In short, this code allows for calculating the nonlinear propagation in a defined fiber by calculating the overlap between the modes and performing propagation of each mode separately with an additional term accounting for the nonlinear index contributions due to all other modes, related to said overlap. As initial conditions we use the temporal profiles shown in Fig.~\ref{fig:2} and we propagate within the same GRIN fiber used in the rest of this work (nonlinear refractive index $n_2 = 3.2\times 10^{-20}\;m^2W^{-1}$).

\begin{figure}[t]
\centering
\includegraphics[width=84mm]{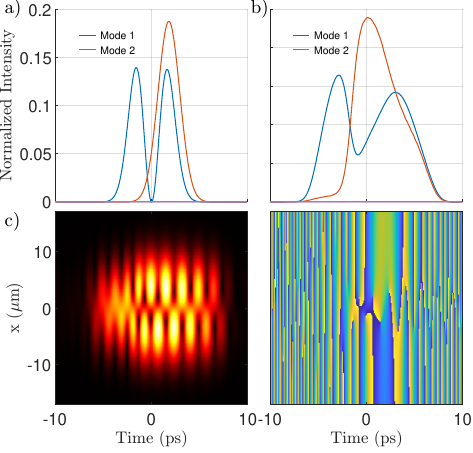}
\caption{Nonlinear propagation of a 500\,fs STOV pulse with $l=1$ in a GRIN fiber at a central wavelength $\lambda_0 = 1030$\,nm. Panel (a) shows the temporal profiles of mode 1 (blue) and mode 2 (red) after 25\,m in the linear regime, and panel (b) in the nonlinear regime when the pulse carries an energy of 0.5 nJ. Panel (b) shows the space-time field amplitude (left) and the phase (right) at $y=0$ after 25\,m of nonlinear propagation.}
\label{fig:6}
\end{figure}

The $l=1$ STOV case provides the most intuitive platform for initial investigations since there are mainly two modes involved. As in the linear regime, the nonlinear regime is much simpler with $l=1$. Therefore we can step back from the space-time picture for the moment and look at the modes interacting in one dimension, where their non-zero overlap allows for interaction in the nonlinear regime purely via the Kerr nonlinearity. In the linear case in Fig.~\ref{fig:6}(a), after 25\,m of propagation the modes have independently spread in time due to their own dispersion and become offset in time due to their different group indices. The fact that they still partially overlap allows for the interference structure seen on the right of Fig.~\ref{fig:5}(a). In the nonlinear case, however, specifically in the case of normal dispersion at $\lambda_0=1030$\,nm, the modes interact through cross-phase modulation.

As seen in Fig.~\ref{fig:6}(b) with 0.5\,nJ of energy for the 500\,fs STOV in the GRIN fiber, the two-lobed temporal structure of mode 1 has trapped mode 2 such that the overlap in time between the two modes is much more significant. Mode 2 is sitting between the temporal lobes of mode 1. This is analogous to what has been seen in other guided-wave nonlinear optics scenarios, often with polarization modes or with separate discrete interacting optical pulses, and has been referred to as domain wall locking~\cite{haelterman94,kockaert99,zhang2010vector,gilles2017polarization,Parra-Rivas:19}. Importantly, in the STOV case as in the other contexts, this does not occur at the zero-dispersion wavelength nor with anomalous dispersion, and only with normal dispersion. The unique aspect here of course, is that we have observed it when pumping with a single space-time object, the STOV, and the domain wall locking is between two modes that make up that initial structure (and have different temporal envelopes). After the nonlinear propagation, the two-lobed temporal structure of mode 1 is degraded, i.e. it no longer has a zero in intensity, due to the influence of the intensity of mode 2. This is in spite of the initially purely real profile of mode 2 on the initially purely imaginary profile of mode 1 (see Fig.~\ref{fig:2}), since the nonlinear interactions are due to the intensity.

After viewing the nonlinear propagation from the purely modal point of view, we can return to the space-time view, shown in Fig.~\ref{fig:6}(c). What we see now is also a result of interference between the modes as in the linear regime, but due to their stronger overlap in time the amplitude has a much more coherent structure---a train of pulses offset from the center, where the minima at $+x$ are the locations of the maxima at $-x$. In fact, at those offset minima, there is a STOV-like structure of $l=\pm1$---a structure strikingly similar the far-field of a chain of STOVs~\cite{huangShunlin24}, but in our case due uniquely to the dispersion and interferences between mode 1 and mode 2. This interference can be understood more deeply when considering the properties of the modes.

Although the dispersion of each mode in the GRIN fiber is also grouped as for absolute phase and group index, the relative difference between the mode groups is so small such that we can consider that the dispersion is approximately equal for mode 1 and mode 2 ($\beta_2$). In general, when two pulses are chirped, i.e. have experienced dispersion, and have a relative time delay, this produces a train of pulses in the amplitude~\cite{rothenberg92}. Such a chirp-and-delay strategy has been used for terahertz generation~\cite{weling94,ahr17,jolly19-2}. In the linear case the temporal offset between mode 1 and mode 2 after propagating a distance $L$ is $\Delta\beta_1^{(2)}L$, and the total chirp accumulated is $\beta_2L$, such that the delay between the individual pulses in the pulse train would be $2\pi\beta_2/\Delta\beta_1^{(2)}\sim1.65$\,ps for our GRIN fiber at 1030\,nm. In fact, we see such interference already appearing with that spacing in the right panel of Fig.~\ref{fig:5}(a) and at the corresponding spacing in the results presented in previous works~\cite{caoQ23,zhangC24}, but in the linear case the overlap in time is poor so it is not so clear. In the nonlinear case as demonstrated here, the overlap is strong due to the locking such that this structure appears at shorter propagation distances and also remains strong and stable during propagation. Although the pulses do not walk-off in the nonlinear case, the inherent frequency conversion that takes place during the nonlinear interaction means that the same relation for the temporal spacing between the pulses in the train holds true, confirmed in Fig.~\ref{fig:6}(c). Finally, the offset in $\pm x$ is a simple result of the odd spatial parity of mode 2.

These results with nonlinear propagation show significant promise. Firstly, they confirm the spatiotemporal extension of well-known nonlinear optical phenomena such as domain-wall locking and dark soliton formation. Therefore the STOV is a relevant platform for exploring fundamental nonlinear optics in multimode guided-wave optics. Secondly, the effect itself of domain-wall locking hints at the possibility of using nonlinear optics to extend the lengths that STOV-like structures are maintained in propagation, or in general enable propagation channels not otherwise possible due to the breakup of the STOV in linear propagation. It has already been demonstrated the technical interest and potential for information transfer using a train of STOVs~\cite{huangShunlin24,liu2024ultrafastburststailoredspatiotemporal}, for example. However, our first nonlinear results do not yet conclusively show an ability to maintain the STOV due to the walk-off of the modes, extra frequency generation, and the degeneration of the singularity in mode 1. Additionally, no such stable structure appears for higher-order STOVs in nonlinear propagation, simply due to the presence of a larger number of spatial modes, as already visible in the right of Fig.~\ref{fig:5}(b) for linear propagation. Regardless, there is vast potential to build on this initial demonstration of nonlinear propagation of a STOV in a multi-mode fiber.

\section{conclusion}

The linear propagation phenomena seen in this work could be viewed as a straightforward result of the modal distribution of a STOV at the input facet and the linear properties of the GRIN (or step-index) fiber, which is why the initial discussion on the non-intuitive modal composition was so important. The general conclusion is that linear propagation will cause the STOV structure to break-up, albeit in different ways and on different length scales depending on the topological charge and central wavelength (and the type of fiber and the resulting modal parameters). This is precisely in opposition to the case of standard spatial optical vortices, where the coupling can be purely to modes within the same group, such that linear propagation does not result in destruction of the singularity (in the absence of environmental perturbations). We discussed importantly the dephasing between modes which, on shorter length scales than where dispersion becomes significant, causes break-up of the STOV structure that is periodic only when the propagation parameters are all grouped together as in the GRIN fiber.

Upon investigating nonlinear propagation, we observed a trapping-like effect due to the domain wall of mode 1 in time locking to that of mode 2. This produced a strong and stable structure of interferences, and hints at a number of interesting directions to be pursued, both fundamental and applied. STOVs are a very unique space-time structure, and the subtleties of their coupling to multimode fibers and subsequent linear and nonlinear propagation are yet to be explored. Especially for nonlinear propagation, we have only shown one specific case of the lowest-order STOV, where higher-order STOVs will surely produce more rich phenomena when properly tamed. This work provides a detailed conceptual foundation, discusses highly multimode fibers and higher-order STOVs, different types of fibers, and numerous aspects of propagation, while presenting the first results of nonlinear propagation. Beyond STOVs, more general space-time structures, potentially also included polarization, will advance the sensing, imaging, and computing applications afforded by multimode waveguides.

\section*{Funding}
Fonds De La Recherche Scientifique - FNRS.

\section*{Disclosures}
The authors declare no conflicts of interest.

\section*{Data availability}
Data underlying the results presented in this paper are not publicly available at this time but may be obtained from the authors upon reasonable request.

\end{document}